\documentclass[fleqn,twoside]{article}
\usepackage{amssymb}
\usepackage{graphicx}
\usepackage{pstricks}
\usepackage{espcrc2}

\newlength{\xs}
\newlength{\ys}
\newlength{\zs}
\newcommand{\w}{\mathrm{w}}

\title{Simulation of a modified Hubbard model with a chemical potential
using a meron-cluster algorithm}

\author{J.C. Osborn\address{Physics Department, University of Utah, Salt Lake City, UT 84112, USA}%
\thanks{This work was done in collaboration with Shailesh Chandrasekharan, J\"urgen Cox and
Uwe-Jens Wiese and was funded by the DOE and NSF.
The simulations were done at Duke partially on computers donated by Intel.}}

\begin{document}

\begin{abstract}
We show how a variant of the Hubbard model can be simulated using a meron-cluster algorithm.
This provides a major improvement in our ability to determine the behavior of these types
of models.
We also present some results that clearly demonstrate the existence of a superconducting state
in this model.
\end{abstract}

\maketitle

Numerical simulations continue to be the most successful way to determine
the low energy behavior of many strongly interacting systems from first principles.
Tremendous advances in simulation methods have been made over the years.
However there are still many problems one can encounter when performing a simulation.
One such problem is the effort required to generate independent configurations.
Many important observables are sensitive to the large scale structure of a configuration
and therefore require require an updating scheme that can make significant changes
to a large fraction of variables.
An important example is observables that are sensitive to topology.
Topology cannot be changed by small, smooth movements and requires updates that
in some sense mimic the large scale fluctuations of the system.

The autocorrelation time quantifies the number of updating steps needed to produce
an independent configuration.
Typically it scales as $\tau \propto \xi^z$ where $\xi$ is a relevant correlation length.
As one approaches a phase transition the correlation length diverges so it is important
to minimize the dynamical exponent $z$ as much as possible.
Standard update algorithms that change only a few variables at a time typically
have $z$ around two.
This can be lowered to $z \approx 1$ or lower
using improved updating schemes such as overrelaxation.
Even with this improvement it is still difficult to make large scale changes.
Cluster algorithms are designed to change a large fraction of the configuration
variables together which greatly reduces the autocorrelations and
can approach $z=0$.

The idea of updating clusters of variables in a Monte Carlo simulation
originated from the work of Swendsen and Wang for the Potts model \cite{SW}.
They used the cluster representation of the Potts model presented earlier by
Fortuin and Kasteleyn \cite{FK} to efficiently update large collections of spins.
This idea was later extended to O(N) spin models by Wolff \cite{UW}.

The reason the algorithm works so well is that the cluster variables
are a natural representation of the long range modes of the system.
Thus cluster algorithms not only provide a means for efficient simulation
but also provide insight to the relevant long distance physics.
The algorithm for classical spin models allows clusters to grow by
joining neighboring sites based on the alignment of the spins.
This can generate clusters of arbitrary shapes which are well suited to
update these models.

Most of the cluster methods for quantum systems are related to the loop algorithm \cite{LA}.
Here the clusters form loops of sites.
This structure can naturally make large changes to the particle worldlines in the path integral.
There have been many improvements made in the updating schemes for loop-clusters \cite{LC}
and it continues to be an area of active research.

Despite the continuing advance of loop algorithms, they are still limited in the types of models
that they can simulate.
Many models with frustration or fermions develop a {\em sign problem} when
written in terms of loop variables.
The sign problem arises from large cancellations of configurations with positive and negative weights.
In frustrated systems of bosons the signs arise due to the competing interactions
while for fermions the signs come directly from the permutation of particle worldlines.
In order to extend loop algorithms to these systems, the clusters must also provide a means
for reducing these cancellations.

The meron-cluster algorithm is a method for solving the sign problem in some models.
It was first introduced to simulate the 2-d O(3) model with a $\theta$-vacuum term \cite{BPW}.
Since then it has been extended to many other models
including certain systems of fermions \cite{Ch00,MCA}.
The fermionic version uses the same cluster variables as in a loop algorithm.
The difference is that these clusters also encode
important information on the fermion sign.
For certain models we are able to make an exact cancellation of all
negative configurations with positive ones leaving
only positive contributions to the partition function.

In this paper we present an overview of some recent work on applying the meron-cluster algorithm
to a variant of the Hubbard model.
The next section contains a brief introduction the standard Hubbard model
and why it is important to many areas of physics.
In section \ref{SECmca} we discuss the key concepts of the meron-cluster
algorithm for spinless fermions.
A complete explanation of the method is available in the literature \cite{Ch00}.
Section \ref{SECmcas} discusses the extension of the algorithm to fermions with spin.
Here we construct a variant of the Hubbard model with the same
symmetries that can be simulated efficiently.
In section \ref{SECef} we discuss the addition of a chemical potential
while maintaining the ergodicity of the algorithm.
Lastly we show numerical results confirming a superconducting
phase in our variant of the Hubbard model.

\section{HUBBARD MODEL}

The Hubbard model was introduced as a simple model for
interacting electrons \cite{Hub}.
Despite its simplicity it is believed to exhibit a wealth of interesting
phenomena of strongly correlated fermions.
It has since been used to describe a variety of systems including BCS
superconductors, high temperature superconductors and
superconductor-insulator and metal-insulator transitions.

The Hamiltonian for this model is
\begin{eqnarray}
H_{hub} &=& -\;t ~ \sum_{<ij>} \sum_{\sigma=\uparrow,\downarrow}
  ~c^{\dagger}_{i,\sigma}\;c^{\phantom{\dagger}}_{j,\sigma} +
   c^{\dagger}_{j,\sigma}\;c^{\phantom{\dagger}}_{i,\sigma}
\nonumber \\
  &+& \sum_{i} ~
  U \left( n_{i,\uparrow} - \frac12 \right)
    \left( n_{i,\downarrow} - \frac12 \right)
\nonumber \\
  & & ~~ -~\mu\; ( n_{i,\uparrow}  + n_{i,\downarrow} )
\label{hm}
\end{eqnarray}
where $<ij>$ stands for all pairs of neighboring sites.
The creation ($c^{\dagger}_{i,\sigma}$) and
annihilation ($c^{\phantom{\dagger}}_{i,\sigma}$) operators for a fermion of spin
$\sigma$ obey the usual anticommutation relations and
$n_{i,\sigma} = c^{\dagger}_{i,\sigma} c^{\phantom{\dagger}}_{i,\sigma}$
is the number operator.
We have written the model such that it remains half-filled (an average
of one particle per site) at zero chemical potential ($\mu$).

The properties of the model depend on the coefficient of the on-site interaction.
In the repulsive model ($U>0$), the interaction can be viewed as the leading
order coulomb repulsion.
Here the model is believed to exhibit properties similar to the high temperature
superconductors.
In particular there is the possibility of finding a transition from antiferromagnetism
to d-wave superconductivity.
This however has not been confirmed numerically due to a severe sign problem \cite{W89}.

The attractive model ($U<0$) serves as a very useful effective model
for interacting fermions.
Here the interaction can originate from the phonon mediated attraction
important to BCS superconductivity.
It is a very general model and can be useful in studying any system of
fermions with an attraction including neutron matter.
The attractive model does not suffer from the sign problem in traditional
simulations as the repulsive model does.
It therefore continues to be a popular subject of numerical simulation.
Despite the extensive studies that have been performed, a precise verification
of a phase transition is still elusive.
This is mainly due to the effort required to simulate at very low temperatures \cite{La96}.

\subsection{Symmetries}

The Hubbard model has two important symmetries that we will want to preserve when
we construct our variant below.
The first is the SU(2)$_s$ spin symmetry defined in terms of the usual spin
operators $S^\alpha = \sum_i \; S_i^\alpha$ with
\begin{eqnarray}
S_i^+ &=& \; c_{i,\uparrow}^\dagger \; c_{i,\downarrow} \nonumber \\
S_i^- &=& \; c_{i,\downarrow}^\dagger \; c_{i,\uparrow} \nonumber \\
S_i^3 &=& \; \frac12 \left( n_{i,\uparrow} - n_{i,\downarrow} \right) ~.
\label{spin}
\end{eqnarray}
When $\mu=0$ there is also an SU(2)$_c$ charge symmetry (also referred to as pseudospin)
whose generators are given by
$J^\alpha = \sum_i \; J_i^\alpha$ with
\begin{eqnarray}
J_i^+ &=& \; (-1)^i~c_{i,\uparrow}^\dagger \; c_{i,\downarrow}^\dagger \nonumber\\
J_i^- &=& \; (-1)^i~c_{i,\downarrow}       \; c_{i,\uparrow}           \nonumber\\
J_i^3 &=& \; \frac12 \left( n_{i,\uparrow} + n_{i,\downarrow} - 1 \right) ~.
\label{pspin}
\end{eqnarray}
The operators $J_i^+$ and $J_i^-$ create and destroy a pair of up and down spin fermions on a site.
The third component $J^3$ is related to the particle number.
The chemical potential couples to this operator and explicitly breaks the SU(2)$_c$
symmetry to the U(1) particle number symmetry.
In two dimensions the particle number symmetry can undergo a
Kosterlitz-Thouless transition to a superconducting state.

\section{\label{SECmca}MERON-CLUSTER ALGORITHM}

The meron-cluster algorithm is explained in detail in the literature \cite{Ch00}.
Here we only present the key concepts necessary to illustrate its
extension to fermions with spin and the inclusion of a chemical potential.

\subsection{Path integral formulation}

For simplicity we will consider only nearest neighbor interactions such that 
the Hamiltonian can be written as $\mathrm{H} = \sum_{<i,j>} h_{i,j}$.
The partition function can be formulated as a path integral using
the Trotter-Suzuki decomposition \cite{TS}.
First the partition function is written as the product of $M$ transfer matrices
\begin{eqnarray}
Z = \mathrm{Tr} \exp( \; -\beta \; \mathrm{H} \; )
  = \mathrm{Tr} \left[ \; \exp( \; -\epsilon \; \mathrm{H} \; ) \; \right]^{\;M}
\end{eqnarray}
with $\epsilon = \beta/M$.
Next the Hamiltonian is split into $2\times d$ parts corresponding to each
interaction direction and whether the interaction goes forward from an even
or an odd site.
The transfer matrix can then be approximated as
\begin{eqnarray}
\exp( \; -\epsilon \; \mathrm{H} \; )
 &\approx& \prod_\mu \exp( \; -\epsilon \sum_{i~\mathrm{even}} h_{i,i+\hat{\mu}} \; ) \nonumber\\
 &\times&  \prod_\mu \exp( \; -\epsilon \sum_{i~\mathrm{odd}}  h_{i,i+\hat{\mu}} \; ) ~.
\end{eqnarray}
Now we introduce an occupation number state between each of the transfer matrix operators.
This gives a total of $N_t = 2d M$ ``timeslices'' each containing a set of occupation numbers.
The individual nearest neighbor interactions that appear between any given neighboring timeslices
all commute and we can thus consider the transfer matrix on each neighboring pair of sites
separately.

\begin{figure}[t]
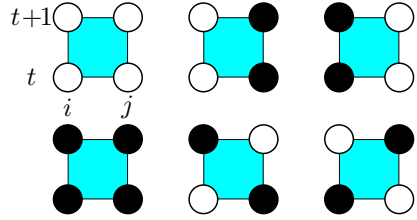

\setlength{\xs}{0.05\textwidth}
\setlength{\ys}{0.05\textwidth}
\setlength{\zs}{0.01\textwidth}
\begin{center}
  \vspace{\ys}
  \begin{tabular}{ccccc}
    \psframe[linewidth=0.1mm,fillstyle=solid,fillcolor=cyan](0,0)(\xs,\ys)
    \pscircle[linewidth=0.2mm,fillstyle=solid,fillcolor=white](0,0){0.2}
    \pscircle[linewidth=0.2mm,fillstyle=solid,fillcolor=white](0,\ys){0.2}
    \pscircle[linewidth=0.2mm,fillstyle=solid,fillcolor=white](\xs,0){0.2}
    \pscircle[linewidth=0.2mm,fillstyle=solid,fillcolor=white](\xs,\ys){0.2}
    \rput{0}(-0.6\xs,0\ys){$t$}
    \rput{0}(-0.6\xs,1\ys){$t\!+\!\!1$}
    \rput{0}(0,-0.5\ys){$i$}
    \rput{0}(\xs,-0.5\ys){$j$}
    \hspace{\xs}
    & \hspace{\zs} &
    \psframe[linewidth=0.1mm,fillstyle=solid,fillcolor=cyan](0,0)(\xs,\ys)
    \pscircle[linewidth=0.2mm,fillstyle=solid,fillcolor=white](0,0){0.2}
    \pscircle[linewidth=0.2mm,fillstyle=solid,fillcolor=white](0,\ys){0.2}
    \pscircle[linewidth=0.2mm,fillstyle=solid,fillcolor=black](\xs,0){0.2}
    \pscircle[linewidth=0.2mm,fillstyle=solid,fillcolor=black](\xs,\ys){0.2}
    \hspace{\xs}
    & \hspace{\zs} &
    \psframe[linewidth=0.1mm,fillstyle=solid,fillcolor=cyan](0,0)(\xs,\ys)
    \pscircle[linewidth=0.2mm,fillstyle=solid,fillcolor=black](0,0){0.2}
    \pscircle[linewidth=0.2mm,fillstyle=solid,fillcolor=black](0,\ys){0.2}
    \pscircle[linewidth=0.2mm,fillstyle=solid,fillcolor=white](\xs,0){0.2}
    \pscircle[linewidth=0.2mm,fillstyle=solid,fillcolor=white](\xs,\ys){0.2}
    \hspace{\xs}
  \end{tabular} \\
  \vspace{1.5\ys}
  \begin{tabular}{ccccc}
    \psframe[linewidth=0.1mm,fillstyle=solid,fillcolor=cyan](0,0)(\xs,\ys)
    \pscircle[linewidth=0.2mm,fillstyle=solid,fillcolor=black](0,0){0.2}
    \pscircle[linewidth=0.2mm,fillstyle=solid,fillcolor=black](0,\ys){0.2}
    \pscircle[linewidth=0.2mm,fillstyle=solid,fillcolor=black](\xs,0){0.2}
    \pscircle[linewidth=0.2mm,fillstyle=solid,fillcolor=black](\xs,\ys){0.2}
    \hspace{\xs}
    & \hspace{\zs} &
    \psframe[linewidth=0.1mm,fillstyle=solid,fillcolor=cyan](0,0)(\xs,\ys)
    \pscircle[linewidth=0.2mm,fillstyle=solid,fillcolor=white](0,0){0.2}
    \pscircle[linewidth=0.2mm,fillstyle=solid,fillcolor=black](0,\ys){0.2}
    \pscircle[linewidth=0.2mm,fillstyle=solid,fillcolor=black](\xs,0){0.2}
    \pscircle[linewidth=0.2mm,fillstyle=solid,fillcolor=white](\xs,\ys){0.2}
    \hspace{\xs}
    & \hspace{\zs} &
    \psframe[linewidth=0.1mm,fillstyle=solid,fillcolor=cyan](0,0)(\xs,\ys)
    \pscircle[linewidth=0.2mm,fillstyle=solid,fillcolor=black](0,0){0.2}
    \pscircle[linewidth=0.2mm,fillstyle=solid,fillcolor=white](0,\ys){0.2}
    \pscircle[linewidth=0.2mm,fillstyle=solid,fillcolor=white](\xs,0){0.2}
    \pscircle[linewidth=0.2mm,fillstyle=solid,fillcolor=black](\xs,\ys){0.2}
    \hspace{\xs}
  \end{tabular}
\end{center}
\vspace{-2\baselineskip}
\caption{Non-zero elements of the plaquette transfer matrix for neighboring sites $i$ and $j$
between timeslices $t$ and $t+1$.}
\label{FIGnztm}
\end{figure}

\subsection{Loop-clusters}

We are now interested in the plaquette transfer matrix
$_{t+1}\!\!\!<\!\!n_i n_j| \exp(-\epsilon h_{ij}) |n_i n_j\!\!>_t$
for the occupation number states of neighboring sites $i$ and $j$ between
timeslices $t$ and $t+1$.
If the interaction conserves particle number then there are only six elements of the transfer
matrix that are non-zero.
These six elements are depicted in Fig.~\ref{FIGnztm}.
The four circles on each plaquette represent occupation numbers with white meaning empty and
black filled.
The two matrix elements where the fermion hops from one site to the other will have a negative sign
if it causes a permutation of the fermion world lines.
We will ignore this extra sign factor for now and will discuss its effect below.

We now introduce a set of bond variables that connect certain sites together.
The connected sites are then constrained to either have the same or opposite occupation numbers.
A {\em flip} of a bond then inverts both occupation numbers it touches.
This way any placement of the bonds can generate several occupation number configurations
through flipping.
The probabilities of placing the different bond variables is chosen such that the 
allowed occupation number configurations are generated with the correct weights
from the transfer matrix.
The model can then be viewed as a system of bonds instead of occupation numbers.

\begin{figure}[t]
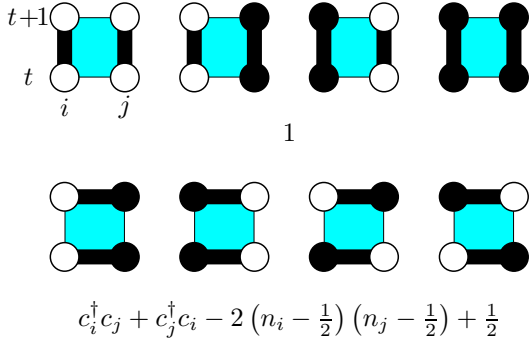

\setlength{\xs}{0.05\textwidth}
\setlength{\ys}{0.05\textwidth}
\setlength{\zs}{0.005\textwidth}
\begin{center}
  \vspace{\ys}
  \begin{tabular}{ccccccc}
    \psframe[linewidth=0.1mm,fillstyle=solid,fillcolor=cyan](0,0)(\xs,\ys)
    \psline[linewidth=2mm](0,0)(0,\ys)
    \psline[linewidth=2mm](\xs,0)(\xs,\ys)
    \pscircle[linewidth=0.2mm,fillstyle=solid,fillcolor=white](0,0){0.2}
    \pscircle[linewidth=0.2mm,fillstyle=solid,fillcolor=white](0,\ys){0.2}
    \pscircle[linewidth=0.2mm,fillstyle=solid,fillcolor=white](\xs,0){0.2}
    \pscircle[linewidth=0.2mm,fillstyle=solid,fillcolor=white](\xs,\ys){0.2}
    \rput{0}(-0.6\xs,0\ys){$t$}
    \rput{0}(-0.6\xs,1\ys){$t\!+\!\!1$}
    \rput{0}(0,-0.5\ys){$i$}
    \rput{0}(\xs,-0.5\ys){$j$}
    \hspace{\xs}
    & \hspace{\zs} &
    \psframe[linewidth=0.1mm,fillstyle=solid,fillcolor=cyan](0,0)(\xs,\ys)
    \psline[linewidth=2mm](0,0)(0,\ys)
    \psline[linewidth=2mm](\xs,0)(\xs,\ys)
    \pscircle[linewidth=0.2mm,fillstyle=solid,fillcolor=white](0,0){0.2}
    \pscircle[linewidth=0.2mm,fillstyle=solid,fillcolor=white](0,\ys){0.2}
    \pscircle[linewidth=0.2mm,fillstyle=solid,fillcolor=black](\xs,0){0.2}
    \pscircle[linewidth=0.2mm,fillstyle=solid,fillcolor=black](\xs,\ys){0.2}
    \hspace{\xs}
    & \hspace{\zs} &
    \psframe[linewidth=0.1mm,fillstyle=solid,fillcolor=cyan](0,0)(\xs,\ys)
    \psline[linewidth=2mm](0,0)(0,\ys)
    \psline[linewidth=2mm](\xs,0)(\xs,\ys)
    \pscircle[linewidth=0.2mm,fillstyle=solid,fillcolor=black](0,0){0.2}
    \pscircle[linewidth=0.2mm,fillstyle=solid,fillcolor=black](0,\ys){0.2}
    \pscircle[linewidth=0.2mm,fillstyle=solid,fillcolor=white](\xs,0){0.2}
    \pscircle[linewidth=0.2mm,fillstyle=solid,fillcolor=white](\xs,\ys){0.2}
    \hspace{\xs}
    & \hspace{\zs} &
    \psframe[linewidth=0.1mm,fillstyle=solid,fillcolor=cyan](0,0)(\xs,\ys)
    \psline[linewidth=2mm](0,0)(0,\ys)
    \psline[linewidth=2mm](\xs,0)(\xs,\ys)
    \pscircle[linewidth=0.2mm,fillstyle=solid,fillcolor=black](0,0){0.2}
    \pscircle[linewidth=0.2mm,fillstyle=solid,fillcolor=black](0,\ys){0.2}
    \pscircle[linewidth=0.2mm,fillstyle=solid,fillcolor=black](\xs,0){0.2}
    \pscircle[linewidth=0.2mm,fillstyle=solid,fillcolor=black](\xs,\ys){0.2}
    \hspace{\xs}
  \end{tabular} \\
  \vspace{\baselineskip}
  $1$
\end{center}
\vspace{\baselineskip}
\begin{center}
  \begin{tabular}{ccccccc}
    \psframe[linewidth=0.1mm,fillstyle=solid,fillcolor=cyan](0,0)(\xs,\xs)
    \psline[linewidth=2mm](0,0)(\xs,0)
    \psline[linewidth=2mm](0,\xs)(\xs,\xs)
    \pscircle[linewidth=0.2mm,fillstyle=solid,fillcolor=white](0,0){0.2}
    \pscircle[linewidth=0.2mm,fillstyle=solid,fillcolor=white](0,\xs){0.2}
    \pscircle[linewidth=0.2mm,fillstyle=solid,fillcolor=black](\xs,0){0.2}
    \pscircle[linewidth=0.2mm,fillstyle=solid,fillcolor=black](\xs,\xs){0.2}
    \hspace{\xs}
    & \hspace{\zs} &
    \psframe[linewidth=0.1mm,fillstyle=solid,fillcolor=cyan](0,0)(\xs,\xs)
    \psline[linewidth=2mm](0,0)(\xs,0)
    \psline[linewidth=2mm](0,\xs)(\xs,\xs)
    \pscircle[linewidth=0.2mm,fillstyle=solid,fillcolor=black](0,0){0.2}
    \pscircle[linewidth=0.2mm,fillstyle=solid,fillcolor=black](0,\xs){0.2}
    \pscircle[linewidth=0.2mm,fillstyle=solid,fillcolor=white](\xs,0){0.2}
    \pscircle[linewidth=0.2mm,fillstyle=solid,fillcolor=white](\xs,\xs){0.2}
    \hspace{\xs}
    & \hspace{\zs} &
    \psframe[linewidth=0.1mm,fillstyle=solid,fillcolor=cyan](0,0)(\xs,\xs)
    \psline[linewidth=2mm](0,0)(\xs,0)
    \psline[linewidth=2mm](0,\xs)(\xs,\xs)
    \pscircle[linewidth=0.2mm,fillstyle=solid,fillcolor=black](0,0){0.2}
    \pscircle[linewidth=0.2mm,fillstyle=solid,fillcolor=white](0,\xs){0.2}
    \pscircle[linewidth=0.2mm,fillstyle=solid,fillcolor=white](\xs,0){0.2}
    \pscircle[linewidth=0.2mm,fillstyle=solid,fillcolor=black](\xs,\xs){0.2}
    \hspace{\xs}
    & \hspace{\zs} &
    \psframe[linewidth=0.1mm,fillstyle=solid,fillcolor=cyan](0,0)(\xs,\xs)
    \psline[linewidth=2mm](0,0)(\xs,0)
    \psline[linewidth=2mm](0,\xs)(\xs,\xs)
    \pscircle[linewidth=0.2mm,fillstyle=solid,fillcolor=white](0,0){0.2}
    \pscircle[linewidth=0.2mm,fillstyle=solid,fillcolor=black](0,\xs){0.2}
    \pscircle[linewidth=0.2mm,fillstyle=solid,fillcolor=black](\xs,0){0.2}
    \pscircle[linewidth=0.2mm,fillstyle=solid,fillcolor=white](\xs,\xs){0.2}
    \hspace{\xs}
  \end{tabular} \\
  \vspace{\baselineskip}
  $c_{i}^{\dagger} c_{j} + c_{j}^{\dagger} c_{i} 
  - 2 \left( n_{i} - \frac12 \right) \left( n_{j} - \frac12 \right) + \frac12 $
\end{center}
\vspace{-2\baselineskip}
\caption{Vertical (upper) and horizontal (lower) bond variables and the
transfer matrix elements they contribute to.
The operator corresponding to each bond variable is also shown.}
\label{FIGbv}
\end{figure}

The simplest set of bond variables one can consider are the two shown in Fig.~\ref{FIGbv}.
The vertical bonds contribute to the four transfer matrix elements shown and
the horizontal bonds also have four contributions.
If we make the approximation $\exp(-\epsilon h_{ij}) \approx 1 - \epsilon h_{ij}$
then it is easy to see what interaction term in the Hamiltonian produces these
bond variables.
The vertical bonds simply correspond to the identity term in the transfer matrix
which is always present.
The horizontal bonds produce the interaction term
\begin{eqnarray}
c_{i}^{\dagger} c_{j} + c_{j}^{\dagger} c_{i} 
  - 2 \left( n_{i} - \frac12 \right) \left( n_{j} - \frac12 \right) + \frac12 ~.
\label{hij}
\end{eqnarray}
Other possible bonds are examined in \cite{Ch00,Ch02}.
Every site is connected to two bonds, one each on the forwards and backwards plaquettes in
the time direction.
The bond variables then form closed loops of sites which do not change
the magnitude of the weight of the configuration when flipped.

\subsection{Meron-clusters}

We now need to consider the previously ignored fermion sign.
Whenever we flip a cluster, the new state will have the same magnitude of weight as the old
but might have a different sign.
Clusters that change the sign of the configuration when flipped are called meron-clusters \cite{BPW}.
The meron-clusters provide a mapping between two different configurations that
cancel and do not contribute to the partition function.

In order to exploit this cancellation we need to satisfy two important conditions.
First the change in sign by flipping any cluster should not depend on how the other clusters
are flipped.
This means that we can cancel the contributions from meron-clusters
independently of the flips of all other clusters.
Second we want to ensure that all negative configurations are canceled with positive
configurations by flipping a meron-cluster.
This ensures that the partition function is the sum of only positive configurations.

We now write the partition function as
\begin{eqnarray}
Z &=& \sum_{[b]}~ W[b]~ \overline{Sign}[b]
\end{eqnarray}
where $W[b]$ is the weight for a bond configuration $b$ and the sum runs over all
possible bond configurations.
If the conditions presented above for canceling meron-clusters hold, then
the average sign is given by
\begin{eqnarray}
\overline{Sign}[b] &=& \prod_{\mathrm{loops}~ \ell}~ \sum_{\rm flips}~ Sign(\ell)
\end{eqnarray}
with the product extending over all loop-clusters for a given set of bonds.
The sum over flips is simply
\begin{eqnarray}
\sum_{\rm flips}~ Sign(\ell) = \left\{
 \begin{array}{cl}
   0 & \mathrm{if~meron-cluster} \\
   2 & \mathrm{otherwise}
 \end{array}
\right. ~.
\end{eqnarray}

As was already stated, configurations with meron-clusters do not contribute to the partition
function.
We can then efficiently simulate the partition function by identifying and avoiding merons.
Typically though, the merons will contribute to observables.
We therefore have to allow some to be generated.
The observables we will consider only get contributions from configurations with zero, one
or two merons.
We can then reject any update that would create any more.

Normally the updates favor the generation of merons and we would end up with mostly
two-meron configurations.
Therefore we must suppress the merons by reweighting configurations with $n$ merons
by a factor of $p^n$ with $p<1$.
We then tune $p$ so that the simulation spends roughly equal time in configurations
of each number of merons.
As the volume gets larger or the temperature is lowered, the factor $p$ must be made
smaller and the tuning becomes more difficult.
However, even for the largest volumes we have run on, this is not a significant
problem.

\section{\label{SECmcas}FERMIONS WITH SPIN}

The simplest way to include spin into the cluster algorithm is to have the clusters only
connect fermions of the same spin and then just duplicate the cluster
configuration for one spin to the other.
If we only use the vertical and horizontal bonds then the Hamiltonian takes the form
$\mathrm{H} = \sum_{<ij>} h_{ij,\uparrow} h_{ij,\downarrow}$
where $h_{ij,\sigma}$ is the operator (\ref{hij}) for the spin $\sigma$ fermions.
This can be rewritten as \cite{Ch02}
\begin{eqnarray}
\sum_{\sigma}~ (c_{i,\sigma}^\dagger c_{j,\sigma} +
                 c_{j,\sigma}^\dagger c_{i,\sigma} )
 \; (n_{ij}-1)(n_{ij}-3)
\nonumber \\
 +~2~ \vec{S}_i \cdot \vec{S}_j~+~2~\vec{J}_i \cdot \vec{J}_j
\nonumber \\
 -~4~ \nu_{i,\uparrow} ~ \nu_{i,\downarrow} ~ \nu_{j,\uparrow} ~ \nu_{j,\downarrow}
\label{vhm}
\end{eqnarray}
with $n_{ij} = n_i + n_j$, $n_i = n_{i,\uparrow} + n_{i,\downarrow}$ and
$\nu_{i,\sigma} = n_{i,\sigma} - 1/2$.
The spin and pseudospin operators are defined in (\ref{spin}) and (\ref{pspin}).

We will study the model (\ref{vhm}) with an on-site interaction and chemical potential
(\ref{hef}) discussed in the next section.
The Hamiltonian for this model is more complicated than the plain Hubbard model
and differs in that it remains strongly interacting at $U=0$.
However it still has the same spin and charge symmetries and
also exhibits a similar phase structure including s-wave superconductivity as we will show below.
The main advantage of this model is that it can be efficiently simulated with the meron-cluster algorithm.

\section{\label{SECef}EXTERNAL FIELDS}

We now want to add the on-site interaction
\begin{eqnarray}
\sum_i ~
 U \left(n_{i,\uparrow}-\frac12\right)\left(n_{i,\downarrow}-\frac12\right)
      ~-~ \mu ~ n_i ~.
\label{hef}
\end{eqnarray}
The simplest method of including this term is to reweight each cluster with
\begin{eqnarray}
\sum_{\rm flips} ~ Sign(\ell)
 &=&   2 \exp(-\frac{\beta}{N_t}\frac14 U S_\ell) \cosh(\beta \mu W_\ell) \nonumber\\
 &\pm& 2 \exp(\frac{\beta}{N_t}\frac14 U S_\ell)
\label{hmw}
\end{eqnarray}
where $S_\ell$ is the total number of sites in a cluster and $W_\ell$ is the temporal winding.
The minus sign is taken if the cluster is a meron and plus otherwise.
For $U\le0$ we can easily see that this weight is always positive and does not introduce any
sign problem.
For $U>0$ the weight can be negative, however if $U>2\mu$ then the negative signs will
always cancel in pairs.
It is likely that in this region the model will be stuck in a charge gap keeping it at half-filling.
It is possible to go to larger $\mu$ before the sign problem becomes severe, but
we have not examined this thoroughly.
We will only consider the attractive model here.

\subsection{Bridge configurations}

The weight (\ref{hmw}) for a cluster can fluctuate to very large values where the simulation can get stuck.
We therefore need a method to avoid this freezing.
The method we employ is based on another application of the meron-cluster algorithm ---
the quantum Heisenberg model in a magnetic field.
An efficient method for simulating this model using meron-clusters was presented in \cite{Ch99}.
Here we will examine the essential features of this algorithm and adapt them to our model.

If we add a magnetic field $h$ in the $z$ direction to the standard loop algorithm for the
quantum Heisenberg model we would then have to reweight each cluster with the factor
\begin{eqnarray}
\exp(\beta h W_\ell/2) + \exp(-\beta h W_\ell/2)
\label{sz}
\end{eqnarray}
where $W_\ell$ is again the temporal winding of the cluster.

\begin{figure}
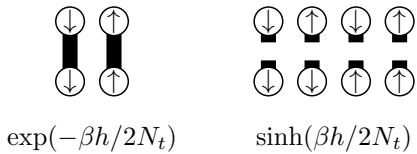

\setlength{\xs}{0.05\textwidth}
\setlength{\ys}{0.05\textwidth}
\setlength{\zs}{0.01\textwidth}
\begin{center}
  \vspace{\ys}
  \begin{tabular}{ccc}
    \begin{tabular}{cc}
      \psline[linewidth=2mm](0,0)(0,\ys)
      \pscircle[linewidth=0.2mm,fillstyle=solid,fillcolor=white](0,0){0.2}
      \pscircle[linewidth=0.2mm,fillstyle=solid,fillcolor=white](0,\ys){0.2}
      \rput{0}(0,0){$\downarrow$}
      \rput{0}(0,\ys){$\downarrow$}
      \hspace{\zs}
      &
      \psline[linewidth=2mm](0,0)(0,\ys)
      \pscircle[linewidth=0.2mm,fillstyle=solid,fillcolor=white](0,0){0.2}
      \pscircle[linewidth=0.2mm,fillstyle=solid,fillcolor=white](0,\ys){0.2}
      \rput{0}(0,0){$\uparrow$}
      \rput{0}(0,\ys){$\uparrow$}
    \end{tabular}
    & \hspace{\zs} &
    \begin{tabular}{cccc}
      \psline[linewidth=2mm](0,0)(0,0.35\ys)
      \psline[linewidth=2mm](0,0.65\ys)(0,\ys)
      \pscircle[linewidth=0.2mm,fillstyle=solid,fillcolor=white](0,0){0.2}
      \pscircle[linewidth=0.2mm,fillstyle=solid,fillcolor=white](0,\ys){0.2}
      \rput{0}(0,0){$\downarrow$}
      \rput{0}(0,\ys){$\downarrow$}
      \hspace{\zs}
      &
      \psline[linewidth=2mm](0,0)(0,0.35\ys)
      \psline[linewidth=2mm](0,0.65\ys)(0,\ys)
      \pscircle[linewidth=0.2mm,fillstyle=solid,fillcolor=white](0,0){0.2}
      \pscircle[linewidth=0.2mm,fillstyle=solid,fillcolor=white](0,\ys){0.2}
      \rput{0}(0,0){$\downarrow$}
      \rput{0}(0,\ys){$\uparrow$}
      \hspace{\zs}
      &
      \psline[linewidth=2mm](0,0)(0,0.35\ys)
      \psline[linewidth=2mm](0,0.65\ys)(0,\ys)
      \pscircle[linewidth=0.2mm,fillstyle=solid,fillcolor=white](0,0){0.2}
      \pscircle[linewidth=0.2mm,fillstyle=solid,fillcolor=white](0,\ys){0.2}
      \rput{0}(0,0){$\uparrow$}
      \rput{0}(0,\ys){$\downarrow$}
      \hspace{\zs}
      &
      \psline[linewidth=2mm](0,0)(0,0.35\ys)
      \psline[linewidth=2mm](0,0.65\ys)(0,\ys)
      \pscircle[linewidth=0.2mm,fillstyle=solid,fillcolor=white](0,0){0.2}
      \pscircle[linewidth=0.2mm,fillstyle=solid,fillcolor=white](0,\ys){0.2}
      \rput{0}(0,0){$\uparrow$}
      \rput{0}(0,\ys){$\uparrow$}
    \end{tabular}
    \\
    \\
    $\exp(-\beta h/2N_t)$ & & $\sinh(\beta h/2N_t)$
  \end{tabular}
\end{center}
\vspace{-2\baselineskip}
\caption{Bond variables and their weights for adding a magnetic field in the $x$ direction to the
quantum Heisenberg model.}
\label{FIGbvsx}
\end{figure}

In \cite{Ch99} they worked around the problem of freezing
by including the magnetic field in the $x$ direction.
This is equivalent to the original problem since the Heisenberg model is symmetric under
rotations.
We can insert this interaction on every site on the lattice.
The transfer matrix for this interaction has the weight $\cosh( \beta h/2N_t )$ if the spin does
not change between the two timeslices and a weight of $\sinh( \beta h/2N_t )$ if it does.

%\begin{eqnarray}
% \left( \begin{array}{cc}
% \cosh( \beta h/2N_t ) & - \sinh( \beta h/2N_t ) \\
% - \sinh( \beta h/2N_t ) & \cosh( \beta h/2N_t )
%\end{array} \right) ~.
%\end{eqnarray}
A set of bond variables that reproduces these weights is shown in Fig.~\ref{FIGbvsx}.
The cluster either continues as normal, or is broken so that the pieces can be flipped independently.
However flipping individual pieces may introduce minus signs into the weight from the
antiferromagnetic interaction \cite{Ch99}.
Pieces of a cluster that start and end on sites of opposite spatial parity will change signs
when flipped.
The sign problem that is introduced is easily solved with a meron-cluster algorithm.
Since the merons are easily identified, one can then suppress their creation and avoid a sign problem.

It is easy to see that this method generates clusters for the partition function
with the same weight as (\ref{sz}).
Consider a cluster with $n$ sites of one spin (e.g. up) and $m$ sites of the opposite spin.
Now add the weights of all magnetic bond configurations that do not produce any merons.
This means that we can put the broken bonds on any combination of even sites or on the odd sites.
Let $\w_s = \exp(-\beta h/2N_t)$ be the weight for a solid bond and $\w_b = \sinh(\beta h/2N_t)$
be for a broken bond.  The total weight is then
\begin{eqnarray}
\w_s^m (\w_s + 2 \w_b)^n + \w_s^n (\w_s + 2 \w_b)^m
\end{eqnarray}
where the extra factor of 2 is due to flipping the different pieces separately.
This is equivalent to (\ref{sz}) with $W_\ell = (n - m)/N_t$.
We now look at why this method is mode efficient.

The key feature of this algorithm is the controlled creation of the merons.
If the merons were never allowed to form the simulation would freeze just like the
plain reweighting method.
The meron configurations then allow the simulation to flow much more easily between configurations
with large weight.
The meron configurations do not contribute to the partition function and are in some sense generated
with an ``unphysical'' weight.
These types of configurations have been referred to as {\it bridges} in the literature \cite{Ib01}.
This naming emphasizes their role in allowing easy passage over areas with small weight.
Of course the merons are important in measuring observables and are thus necessary anyway.
In this example they have an added feature of avoiding freezing.

Just as we summed the weights for the zero-meron sector, we can now see exactly what weight the
meron configurations are generated with.
In this algorithm the meron pieces are always produced in pairs.
The weight of the two-meron configuration is given by summing over all ways of putting the broken
bonds such that there are only two pieces that have their ends on one odd and one even site.
This means that we can split the cluster into two parts and one part has broken bonds only
on the odd sites and the other part has breaks only on the even sites.
Given a splitting of the cluster such that one side has $n_e$ even sites and the other has $m_o$ odd
sites, the weight of that splitting is
\begin{eqnarray}
\w_s^l + \w_s^{l-n_e-m_o} (\w_s + 2 \w_b)^{n_e} (\w_s + 2 \w_b)^{m_o}
\label{2mw}
\end{eqnarray}
where $l$ is the length of the cluster.
The full weight of the two-meron sector is the sum of (\ref{2mw}) over all splittings while avoiding
possible duplicates.
This weight is not as easy to calculate as in the zero-meron sector.
We can estimate its magnitude by identifying the leading term for large fields.
The leading term of (\ref{2mw}) is
\begin{eqnarray}
\exp(\beta h[n_e-n_o+m_o-m_e]/2N_t)
\end{eqnarray}
where we have made the substitution $l = n_e + n_o + m_e + m_o$ with $n_o$ the number of odd sites
on the same side of the splitting as $n_e$ and $m_e$ is the number of
even sites on the side with $m_o$.
The terms $w_n = n_e-n_o$ and $w_m = m_e-m_o$ give the temporal extent of each part of the splitting.
The total winding number of a cluster is given by $W_\ell = (w_n+w_m)/N_t$ and is what appears in
the zero-meron weight (\ref{sz}).
Instead the two-meron weight is related to the maximum difference in temporal winding
of the two parts when the cluster is split.

Consider now why the traditional reweighting scheme fails.
If a cluster with a large positive winding is next to a cluster of large negative winding
the two can never join.
This is because the cancellation of the positive and negative windings would produce a cluster
of much smaller weight.
However if this combined cluster were given the weight for the two-meron configuration the weight
would typically be greater than that of the two separate clusters.
The clusters would now be allowed to join.

Instead of implementing the meron-cluster algorithm for this model we can just generate 
configurations using either the regular (zero-meron) weight or the bridge (two-meron) weight.
For the Hubbard model we need to reweight clusters according to (\ref{hmw}).
For simplicity we then choose the bridge configuration weight to be
\begin{eqnarray}
2 \exp(-\frac{\beta}{N_t}\frac14 U S_\ell) \cosh(\beta \mu [w_n-w_m]/N_t)
\end{eqnarray}
with $w_n$ and $w_m$ defined as above for the splitting that maximizes this quantity.
This weight has the same behavior for large chemical potential as the true two-meron configurations
would and can be calculated efficiently in the update process.

As with the implementation of the meron-cluster algorithm the simulation will want to produce
mostly bridge configurations.
We therefore need to suppress the bridge configurations by some constant factor.
For large fields choosing the right factor becomes difficult, however for the results
presented below we find good movement between the regular and bridge configurations.

\section{NUMERICAL RESULTS}

\begin{figure}[t]
\includegraphics*[width=0.46\textwidth]{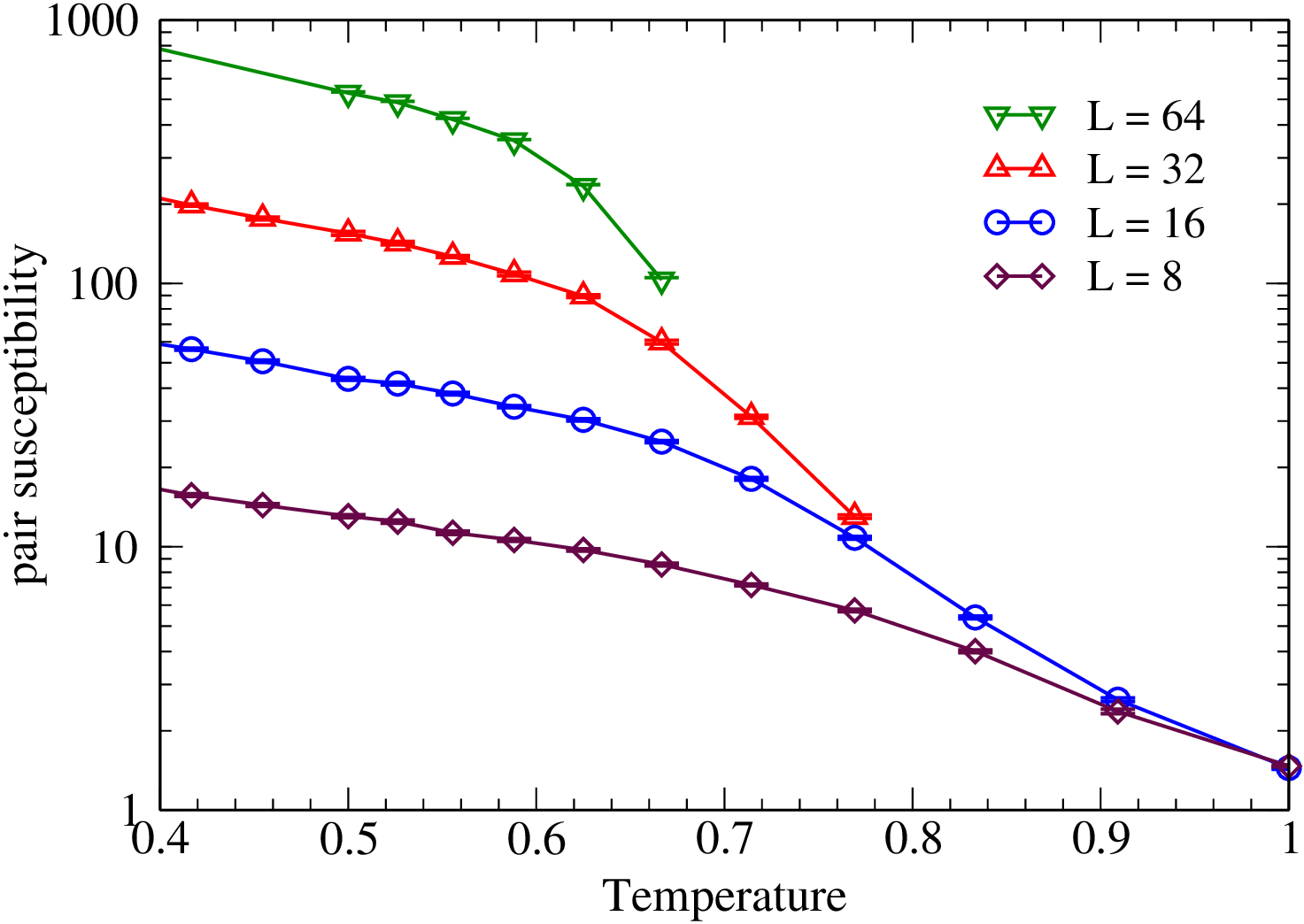}
\vspace{-2\baselineskip}
\caption{Pair susceptibility versus temperature for several system sizes.}
\label{FIGstm1}
\end{figure}

We simulated the model (\ref{vhm}) with an chemical potential of $\mu=1$ in two spatial dimensions.
All results shown are with $U=0$.
The model is still strongly interacting here and has a qualitatively similar behavior as for $U<0$.
To make sure that the simulations weren't getting stuck in any area of phase space
we ran at least ten independent simulations for each set of parameters.
Each run used $10^4$ configurations for the measurements.
The error bars shown are estimated from the variance over the independent runs.

In two dimensions long range order is forbidden according to the Mermin-Wagner theorem.
However long range correlations can emerge through a Kosterlitz-Thouless phenomenon \cite{KT}.
This can be observed by measuring the correlations of the pair annihilation operator
$\Delta = \sum_i c_{i,\downarrow} c_{i,\uparrow}$ with the
pair creation operator $\Delta^\dagger$.
The scaling behavior of the pair correlations is most easily observed through the
pair susceptibility which we define as
\begin{eqnarray}
\chi_L =
 \frac2{Z V} \int_0^\beta \mathrm{d}t ~
 \mathrm{Tr} \left[ \mathrm{e}^{-(\beta-t) \mathrm{H}} ~ \Delta^\dagger ~
 \mathrm{e}^{-t \mathrm{H}} ~ \Delta \right] .
\end{eqnarray}
In a finite volume for $T<T_c$ the pair susceptibility should scale with the
system size ($L$) as
\begin{eqnarray}
\chi_L \; \propto \; L^{\gamma(T)}
\end{eqnarray}
with $\gamma(T_c)=7/4$ and $\gamma(T)$ approaches 2 as $T$ approaches zero.
Above $T_c$ the susceptibility will approach a constant.

\begin{figure}[t]
\includegraphics*[width=0.46\textwidth]{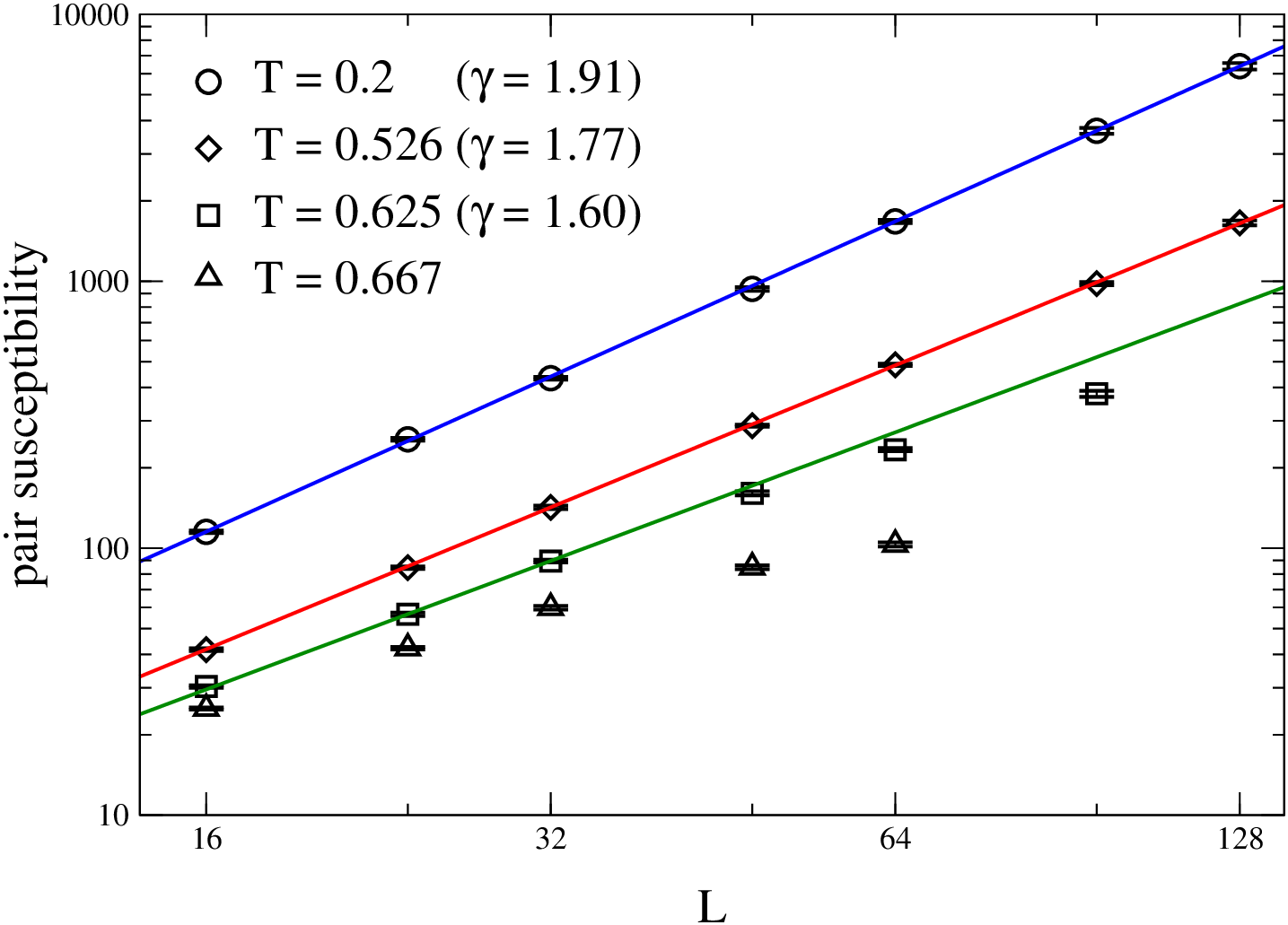}
\vspace{-2\baselineskip}
\caption{Pair susceptibility versus system size for several temperatures.}
\label{FIGslm1}
\end{figure}

In Fig.~\ref{FIGstm1} we show the pair susceptibility versus temperature for several system sizes.
At high temperatures the susceptibility does not grow with the system size.
As the temperature is lowered the susceptibility for large volumes grows rapidly.
For low enough temperatures the susceptibility appears to scale as a power of the volume.

To see this scaling more clearly, in Fig.~\ref{FIGslm1} we show the
pair susceptibility versus system size on a log-log scale.
At the two lowest temperatures in the plot ($T=0.2$ and $T=0.526$) we find excellent
power law scaling with exponents of $\gamma=1.91$ and $\gamma=1.77$ respectively.
The $\gamma$ of 1.77 is just above the prediction at $T_c$.
This suggests that $T_c$ is just above 0.526.
Indeed as we raise the temperature to 0.625 we no longer find power law scaling.
Smaller volumes ($L\le32$) appear to scale, but when we go to larger volumes
we see that this is clearly not the case.
It is therefore crucial to simulate on large volumes to accurately determine
the critical behavior of the system.
Conventional methods are limited to much smaller lattice sizes and therefore
have not been able to accurately measure any long range pair correlations in the
attractive Hubbard model.

\begin{figure}[t]
\includegraphics*[width=0.46\textwidth]{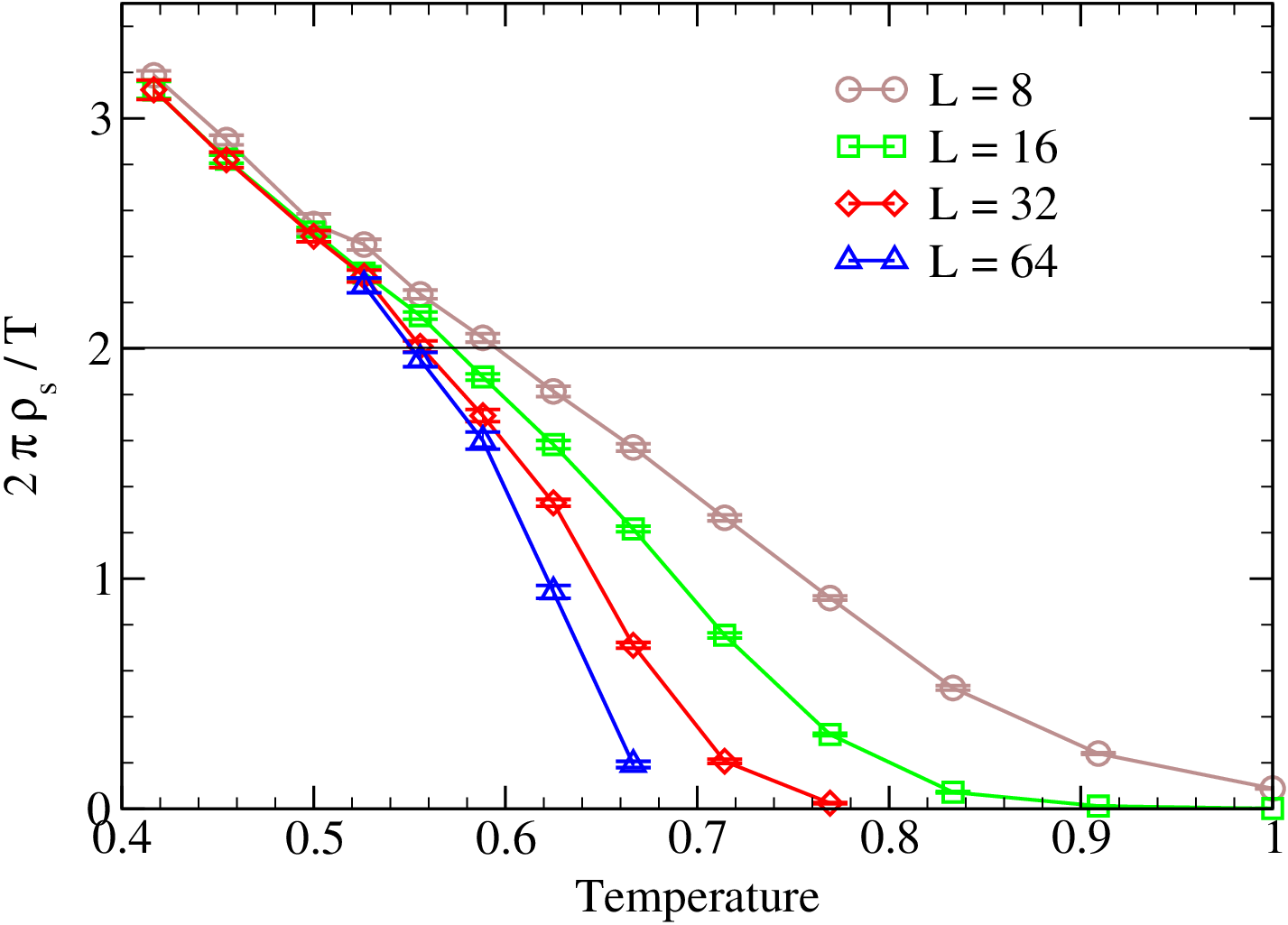}
\vspace{-2\baselineskip}
\caption{Scaled winding number versus temperature for several system sizes.}
\label{FIGwtm1}
\end{figure}

Another important observable for studying superconductivity is the
superfluid density.
It can be defined in terms of the spatial winding as \cite{SD}
\begin{eqnarray}
\rho_s &=& \frac{T}{4} \left< \; (W_x/2)^2 + (W_y/2)^2 \; \right>
\end{eqnarray}
where $W_x$ ($W_y$) is the total number of particles winding around the
boundary in the x (y) direction.
This quantity is sensitive to the phase coherence of the system and is
thus used to indicate superconductivity.

The finite size scaling formula for $T<T_c$ is given by \cite{Ha98}
\begin{eqnarray}
\frac{2\pi}{T}\rho_s = 2 + \sqrt{A}~ \coth(\sqrt{A}~ \log(L/L_0) )
\label{wns}
\end{eqnarray}
with $A=0$ at $T=T_c$.
For $T>T_c$ the superfluid density vanishes at large volumes, and jumps to
a universal value at $T_c$.

In Fig.~\ref{FIGwtm1} we show the scaled superfluid density for different volumes
as a function of temperature.
At large temperatures we can clearly see that it goes to zero as the volume
is increased.
For low temperatures we do not see any scaling with the volume.
At $T=0.526$ there is little scaling with the volume and a fit to (\ref{wns})
verifies that it approaches a constant at infinite volume.
This is consistent with the conclusion reached from the pair susceptibility that
this temperature is below $T_c$.
However at $T=0.555$ there is a noticeable scaling and the fit predicts that the
superfluid density should scale to zero.

\begin{figure}[t]
\includegraphics*[width=0.46\textwidth]{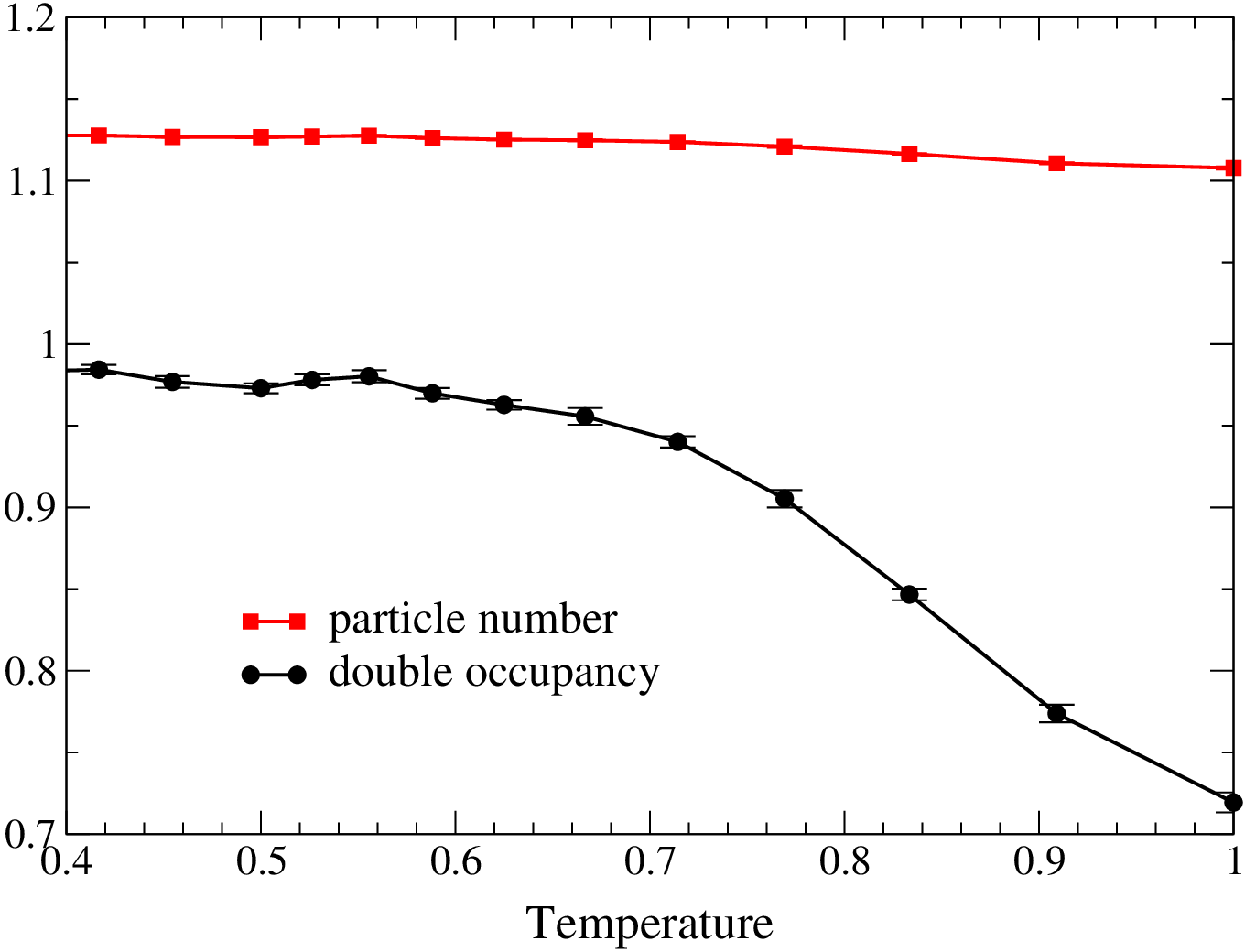}
\vspace{-2\baselineskip}
\caption{Particle number and double occupancy versus temperature.}
\label{FIGdtm1}
\end{figure}

To show that we are simulating a fermionic model we can look at the double
occupancy which is twice the average number of sites that have both an up
and down spin fermion.
If the fermions are all tightly bound in on-site pairs then this quantity
would be equal to the total particle number.
In Fig.~\ref{FIGdtm1} we see that the average number of particles stays roughly
constant just above 1.1 as the temperature is changed.
The double occupancy, however, starts off fairly low at high temperatures and
rapidly increases as $T_c$ is approached.
Thus we are able to observe the formation of the pairs responsible for superconductivity
and not just simulate the Bose-Einstein condensation of preformed boson pairs.

\section{CONCLUSIONS}

We have shown how a meron-cluster algorithm can be used to efficiently simulate a
variant of the Hubbard model on large lattices.
This has allowed us to accurately study the behavior of this model near the
superconducting phase transition.
Currently the meron-cluster algorithm is limited to specific models.
However, just as cluster algorithms for bosonic systems have flourished over the past
ten years,
hopefully fermionic cluster algorithms will too.

We are currently using the model shown here to examine the superconductor-insulator
transition in the presence of disorder.
Preliminary results suggest the possibility of a localized superconducting state,
though further work is needed to determine the precise nature of this state.
Another future interest is extending the fermion content to include isospin or
color degrees of freedom.
This may provide some precise results for new and interesting types of models.


\begin{thebibliography}{9}
\bibitem{SW}  R.H. Swendsen and J.-S. Wang, Phys. Rev. Lett. 58 (1987) 86.
\bibitem{FK}  P.W. Kasteleyn and C.M. Fortuin, J. Phys. Soc. Jpn. Suppl. 26 (1969) 11;
              C.M. Fortuin and P.W. Kasteleyn, Physica 57 (1972) 536.
\bibitem{UW}  U. Wolff, Phys. Rev. Lett. 62 (1989) 361.
\bibitem{LA}  H.G. Evertz, G. Lana and M. Marcu, Phys. Rev. Lett. 70 (1993) 875.
\bibitem{LC}  For a review see H.G. Evertz, in
              {\em Numerical Methods for Lattice Quantum Many-Body Problems},
              ed. D.J. Scalapino, Perseus books, Frontiers in Physics (2000).
\bibitem{BPW} W. Bietenholz, A. Pochinsky and U.-J. Wiese, Phys. Rev. Lett. 75 (1995) 4524.
\bibitem{Ch00}
S. Chandrasekharan, J. Cox, K. Holland and U.-J. Wiese, Nucl. Phys. B 576 (2000) 481.
\bibitem{MCA}
S. Chandrasekharan, Nucl. Phys. B (Proc. Suppl.) 83 (2000) 774;
J. Cox, C. Gattringer, K. Holland, B. Scarlet, U.-J. Wiese, Nucl. Phys. B (Proc. Suppl.) 83 (2000) 777;
S. Chandrasekharan and J.C. Osborn, Phys. Lett. B 496 (2000) 122;
S. Chandrasekharan and J.C. Osborn, arXiv:cond-mat/0109424.
\bibitem{Hub} J. Hubbard, Proc. Roy. Soc. A 276 (1963) 238.
\bibitem{W89}
S.R. White, D.J. Scalapino, R.L. Sugar, E.Y. Loh, J.E. Gubernatis and
R.T. Scalettar, Phys. Rev. B 40 (1989) 506.
\bibitem{La96} R. Lacaze, A. Morel, B. Petersson and J. Schr\"oper, Eur. Phys. J. B2 (1998) 509.
\bibitem{TS} H.F. Trotter, Proc. Am. Math. Soc. 10 (1959) 545; M. Suzuki, Prog. Theor. Phys. 56 (1976) 1454.
\bibitem{Ch02} S. Chandrasekharan, J. Cox, J.C. Osborn and U.J. Wiese, arXiv:cond-mat/0201360.
\bibitem{Ch99} S. Chandrasekharan, B. Scarlet and U.J. Wiese, arXiv:cond-mat/9909451.
\bibitem{Ib01} See e.g. Y. Iba, Int. J. Mod. Phys. C 12 (2001) 623.
\bibitem{KT}
J.M. Kosterlitz and D.J. Thouless, J. Phys. C 6 (1973) 1181;
J.M. Kosterlitz, J. Phys. C 7 (1974) 1046.
\bibitem{SD}
E.L. Pollock and D.M. Ceperley, Phys. Rev. B 36 (1987) 8343;
H. Weber and P. Minnhagen, Phys. Rev. B 37 (1988) 5986.
\bibitem{Ha98} 
K. Harada and N. Kawashima, J. Phys. Soc. Jpn. 67 (1998) 2768.
\end{thebibliography}
\end{document}